\documentstyle[11pt,newpasp,twoside,epsf]{article}
\def\edcomment#1{\iffalse\marginpar{\raggedright\sl#1\/}\else\relax\fi}
\reversemarginpar

\begin{document}
\def\lapp{\ifmmode\stackrel{<}{_{\sim}}\else$\stackrel{<}{_{\sim}}$\fi}
\def\gapp{\ifmmode\stackrel{>}{_{\sim}}\else$\stackrel{>}{_{\sim}}$\fi}
\title{New limits on the strong equivalence principle from two
long-period circular-orbit binary pulsars}
\author{D.R. Lorimer,$^{1}$ P.C.C. Freire$^{2}$}
\affil{$^{1}$University of Manchester, Jodrell Bank Observatory, SK11 9DL, UK\\
$^{2}$NAIC, Arecibo Observatory, HC3 Box 53995, Arecibo, PR 00612, USA}

\begin{abstract}
Following a brief review of the principles of the strong equivalence
principle (SEP) and tests for its violation in the strong and weak
gravitational field regimes, we present preliminary results of new
tests using two long-period binary pulsars: J0407+1607 and J2016+1947.
PSR J0407+1607 is in a 669-day orbit around a $\gapp 0.2$ M$_{\odot}$
companion, while J2016+1947 is in a 635-day orbit around a $\gapp 0.3$
M$_{\odot}$ companion. The small eccentricities of both orbits 
($e \sim 10^{-3}$)
mean that these systems reduce previous limits on SEP violation 
by more than a factor of 4.
\end{abstract}

\section{Equivalence principles and gravitational self-energy}

The principle of equivalence between gravitational force and
acceleration is a common feature to all viable theories of
gravity. The Strong Equivalence Principle (SEP), however, is unique to
Einstein's general theory of relativity (GR).  Unlike the weak
equivalence principle (which dates back to Galileo's demonstration
that all matter free falls in the same way) and the Einstein
equivalence principle from special relativity (which states that the
result of a non-gravitational experiment is independent of rest-frame
velocity and location), the SEP states that free fall of a body is
completely independent of its gravitational self energy.

Before examining how the SEP can be tested, let us first review the
gravitational self energy, $\epsilon$, which is a useful
quantity for distinguishing between strong or weak gravitational
fields.  Expressed in terms of the rest-mass energy of a body of mass
$M$ and size $R$, $\epsilon = -GM/Rc^2$. For most bodies, $\epsilon$
is vanishingly small. For example $\epsilon_{\rm human} \sim
-10^{-26}$, $\epsilon_{\rm Earth} \sim -5 \times 10^{-10}$ and even
$\epsilon_{\odot} = -2 \times 10^{-6}$. Only for compact objects does
$\epsilon$ become significant and we enter the ``strong-field''
regime.  For a white dwarf $\epsilon_{\rm WD} \sim -10^{-4}$, for a neutron
star $\epsilon_{\rm NS} \sim -0.3$ and for a non-rotating black hole,
$\epsilon_{\rm BH}=-0.5$.

\section{Testing the strong equivalence principle}

If the SEP is violated, then the ratio of inertial mass to
gravitational mass of a test particle differs from unity by an amount
\begin{equation}
\Delta = \eta \epsilon + \eta' \epsilon^2 + \cdots,
\end{equation} 
where $\eta$ parameterises the violation in terms of $\epsilon$.
A violation of the SEP would mean that two bodies of unequal
self-energies fall differently in an external gravitational field. 
If these bodies were in orbit about one another, their differential free
fall would cause the orbit to be ``polarized'' in the direction of the
external field. This so-called ``Nordvedt effect'' was first proposed
as a test of the SEP for the Earth-Moon system in the Sun's
gravitational field (Nordvedt 1968a,b).

While lunar laser ranging measurements of the Nordvedt effect
place fairly stringent constraints ($|\eta|<0.0016$; see Will
2001 for a review), since $\epsilon_{\rm Earth}$
and $\epsilon_{\rm Moon}$ are so small, this test only
applies to the weak-field regime. 

Damour \& Sch\"afer (1991) proposed a further means to
test the SEP by an analagy of the Nordvedt effect on Galactic neutron
star-white dwarf binary systems. In this case, the external
gravitational field is provided by the Galaxy rather than the Sun and,
as shown in Fig.~1, for a violation of the SEP polarizes the eccentricity
in the direction of the local gravitational field. Due
to the larger and significantly different self energies of neutron
stars and white dwarfs over solar system bodies, these binaries
test the SEP in the strong-field regime. 

\begin{figure}[h!]
\plotfiddle{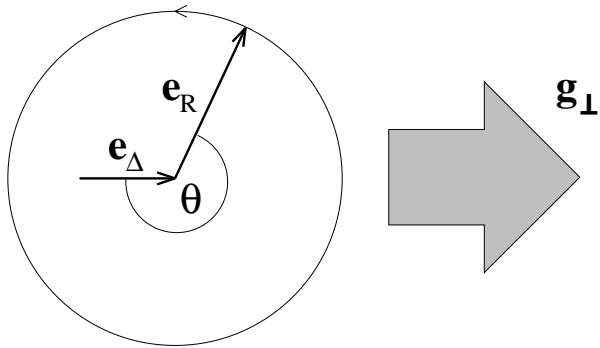}{4cm}{-90}{50}{50}{-100}{130}
\caption{Diagram from Wex (1997) showing two competing effects on
the orientation of a binary system's eccentricity: the
component in the direction of the local gravitational field due to an
SEP violation ($e_{\Delta}$) and the rotation due to relativistic periastron
advance ($e_R$). }
\end{figure}

We shall leave it as an exercise to the
interested reader (see Damour \& Sch\"afer 1991 for details) to show
that the constraint provided by a binary system
\begin{equation}
\Delta \leq \frac{8 \pi^2 G M e \xi(\theta)}
{gc^2 P_b^2 \left[1 - (\cos i \cos \lambda + \sin i 
\sin \lambda \sin \Omega)^2 \right]^{1/2}}.
\end{equation}
Here $M$ is the total mass of the system, $e$ is the orbital
eccentricity, $P_b$ is the orbital period, $i$ is the inclination
angle between the plane of the orbit and the plane of the sky,
$\lambda$ is the direction between the gravitational field
and the line of sight, $\Omega$ is the longitude of the ascending node and
the geometrical factor
\begin{equation}
\xi(\theta) = \left\{ 
    \begin{array}{lll}
      1/\sin(\theta) & \mbox{for} & 0      < \theta \leq \pi/2\\
      1              & \mbox{for} & \pi/2  < \theta < 3\pi/2\\
     -1/\sin(\theta) & \mbox{for} & 3\pi/2 \leq \theta \leq 2\pi
    \end{array} 
 \right.
\end{equation}
where $\theta$ is the angle shown in Fig.~1. From equation (2) is is clear that 
binary systems with large values of $P_b^2/e$, i.e.~long-period
circular orbit systems, are the best systems for placing
limits on $\Delta$. In order for the above equation (3) to
be valid, two conditions must be met:
\begin{itemize}
\item The relativistic periastron advance must have completed
sufficient revolutions so that its effects on $e_{\Delta}$
can be considered to have averaged out. Following Wex (1997),
we may express this in terms of an age constraint:
\begin{equation}
  t_{\rm system} > 2 \times 10^8 \,{\rm yr}\, 
  \left(\frac{P_b}{10^3 {\rm days}}\right)^{5/3}
  \left(\frac{M}{M_{\odot}}\right)^{-2/3}.
\end{equation}
\item The relativistic periastron advance must be significantly
larger than the Galactic rotational velocity. 
\end{itemize}

Damour \& Sch\"afer (1991) used two pulsars
that satisfied these constraints and calculated $\Delta$ using a
simple Monte Carlo simulation to randomize over the unknown
angles $\theta$ and $\Omega$. Their analysis was significantly
revised by Wex (1997) using an ensemble of seven binaries.
Wex's analysis is summarized in Table 1. For each
system, we show the probability of $\Delta$ exceeding a given
fractional level between 1 and 0.2\%. In other words, the
entries give the probability of the SEP being violated
at the given level. In terms of the simulation, these numbers
are readily calulated by simply recording the number of times
$\Delta$ exceeds a given level and dividing this number by
the total number of trials.

\begin{table}[h]
\begin{center}
\begin{tabular}{l|c|c|c|c|c}
\hline
\hline
Pulsar &     1.0\% & 0.5\%& 0.4\%& 0.3\%& 0.2\% \\
\hline
\hline
J1455--3330 & 0.18 & 0.68 & 1.00 & 1.00 & 1.00\\
J1640+2224  & 0.23 & 0.63 & 0.92 & 1.00 & 1.00\\
J1643--1224 & 0.27 & 0.60 & 0.77 & 1.00 & 1.00\\
J1713+0747  & 0.08 & 0.18 & 0.23 & 0.39 & 1.00\\
B1953+29    & 0.26 & 0.56 & 1.00 & 1.00 & 1.00\\
J2019+2425  & 0.20 & 0.45 & 0.57 & 0.79 & 1.00\\
J2229+2643  & 0.17 & 0.76 & 1.00 & 1.00 & 1.00\\
\hline
\hline
Product        & 0.00 & 0.01 & 0.09 & 0.31 & 1.00\\
1 -- Product   & 1.00 & 0.99 & 0.91 & 0.69 & 0.00\\
\hline
\hline
\end{tabular}
\end{center}
\caption{Results of Wex's (1997) analysis displayed in a slightly
different form to the original version. For each binary system
we display the probability that $\Delta$ exceeds a given
fractional level listed at the top of each column. The products
of these probabilities are given to two decimal places (see text).}
\end{table}

Since each binary provides an independent test of the SEP, Wex
was able to place strong constraints by taking the combined
probabilities for each systems (i.e.~the product of the individual
values) the ``1--product'' entry listed in the table gives the
combined probability that the SEP is valid at the given fractional
level. From this table, we conclude at the 90\% confidence level
that $\Delta < 0.004$. As smaller fractional
limits on $\Delta$ are probed it can be seen that each binary
system begins to ``drop out'' of the joint test when the
individual probability reaches unity. As a result,
these systems place no constraints on the SEP at the level $\Delta<0.002$.

\section{Two new long-period binary systems}

As discussed throughout this meeting, there are many exciting results
emerging from the flood of binary pulsars discovered in recent
years. As far as the SEP is concerned, most of these systems do not
improve upon Wex's (1997) analysis due to either small ages, or more
commonly, their low values of $P_b^2/e$. Two exceptions, however, are
PSRs J0407+1607 (Lorimer et al.~in preparation) and J2016+1947
(Navarro, Anderson \& Freire 2003). Both of these
binary pulsars were discovered in 430-MHz Arecibo surveys during the
1990s and have only recently been observed long enough to 
accurately determinatine their orbital parameters and to establish
that they are old enough to satisfy equation (4).

\begin{table}[h]
\begin{center}
\begin{tabular}{l|c|l|l}
\hline
\hline
Parameter & Unit & PSR J0407+1607 & PSR J2016+1947 \\
\hline
\hline
Spin period &  ms &   25.7      &     64.9           \\
Characteristic age & Gyr & $>$1    &       2.5         \\
Orbital period & days&   669       &   635          \\
Eccentricity   &     &  0.00095    &   0.00148         \\
$P_b^2/e$      & days$^2$ & $4.7\times 10^8$ & $2.7\times 10^8$     \\
Companion mass &  M$_{\odot}$& $0.19/\sin i$& $0.29/\sin i$\\
\hline
\end{tabular}
\end{center}
\caption{Spin and orbital parameters for the two newly-discovered 
binary pulsars relevant to the SEP tests. The companion
masses are based on the Keplerian orbital parameters 
in terms of the unknown orbital inclination $i$ and
assuming a pulsar mass of 1.35 M$_{\odot}$.}
\end{table}

As can be seen from Table 2, the two new pulsars
have very large values of $P_b^2/e$ and greatly
exceed the previous best system J1713+0747 ($6\times 10^7$ days$^2$)
used by Wex (1997). We note in passing that both these
systems fall on the orbital period-eccentricity
relation predicted by Phinney (1992).

\section{New limits on the SEP}

To demonstrate the level at which the new systems improve
the constraints on $\Delta$, in Table 3 we present some preliminary
results of a new simulation which closely follows Wex's
(1997) analysis. As in this earlier analysis we exclude the long-period binary
pulsars B0820+02 and B1800--27 since it is not clear
whether they satisfy the age constraint.

For most of the pulsars under consideration,
as expected, there is no contribution to each fractional
value of $\Delta$ and the test is essentially dominated
by the two new pulsars for the regime $\Delta \leq 0.0002$.
As a firm upper limit, we find $\Delta < 0.003$.
At the 90\% confidence level, we conclude that $\Delta < 0.0009$.
This represents an improvement by a factor of more than four 
over the previous limits.

\begin{table}[h]
\begin{center}
\begin{tabular}{l|c|c|c|c|c}
\hline
\hline
Pulsar      & 0.3\%& 0.2\%&0.09\%&0.08\%&0.05\%\\
\hline
\hline
J1455--3330 & 1.00 & 1.00 & 1.00 & 1.00 & 1.00\\
J1640+2224  & 1.00 & 1.00 & 1.00 & 1.00 & 1.00\\
J1643--1224 & 1.00 & 1.00 & 1.00 & 1.00 & 1.00\\
J1713+0747  & 0.39 & 1.00 & 1.00 & 1.00 & 1.00\\
B1953+29    & 1.00 & 1.00 & 1.00 & 1.00 & 1.00\\
J2019+2425  & 0.79 & 1.00 & 1.00 & 1.00 & 1.00\\
J2229+2643  & 1.00 & 1.00 & 1.00 & 1.00 & 1.00\\
\hline
J0407+1607  & 0.06 & 0.09 & 0.21 & 0.24 & 0.46\\
J2016+1957  & 0.11 & 0.17 & 0.47 & 0.56 & 0.92\\
\hline
\hline
Product     & 0.00 & 0.02 & 0.10 & 0.13 & 0.42\\
1 -- Product& 1.00 & 0.98 & 0.90 & 0.87 & 0.58\\
\hline
\hline
\end{tabular}
\end{center}
\caption{Preliminary results showing the tests of violations
of the SEP for various fractional values of the parameter $\Delta$.
As in Table 1, we list the products of the individual
probabilities for each system. The ``1--Product'' entry represents
the probability that the SEP is valid at a given fractional level.}
\end{table}

\section{Implications}

The greater self energy of neutron stars and white dwarfs over
the Earth and Moon used in the solar system tests allow us to
probe the term of order $\epsilon^2$ in equation (1). Assuming
from the lunar laser ranging experiments that $\eta$ is negligible
in this expression, and ignoring the small self energy contribution
from the white dwarf, we may write
\begin{equation}
	\Delta \simeq \epsilon_{\rm NS}^2 (\varepsilon/2 + \zeta),
\end{equation}
where the parameters $\varepsilon$ and $\zeta$ are zero in GR but
non-zero in alternative scalar-tensor theories (e.g.~Damour \& Esposito-Farese
1995). Assuming that $\epsilon_{\rm NS}=0.3$,
our preliminary results place a 90\% confidence limit on the sum
\begin{equation}
	| \varepsilon/2 + \zeta | < 0.001.
\end{equation}

\end{document}